\documentclass[10pt,oneside]{amsart}

\usepackage{tikz}
\usetikzlibrary{automata, positioning, arrows}

      \usepackage{amssymb}
\usepackage{amsfonts,amstext}
\usepackage{graphicx}
\usepackage{url}
\usepackage{amsmath,amstext,amssymb,amscd}
\usepackage{verbatim} 
\usepackage{mathtools}
      \usepackage{tabularx,booktabs}
      \usepackage{multicol}

\usepackage{amssymb,latexsym}
\usepackage{amsmath}
\usepackage{amsthm}
\usepackage{amssymb}
\usepackage{makecell}
\newtheorem{thm}{Theorem}[section]

\theoremstyle{remark}

\theoremstyle{definition}
\newtheorem{mydef}[thm]{Definition}

\usepackage{amsthm}
\allowdisplaybreaks

\usepackage{multicol}

      \makeatletter
      \def\@setcopyright{}
      \def\serieslogo@{}
     
      \makeatother

\begin{document}

\author{Adam Graham-Squire}
\address{Adam Graham-Squire, Department of Mathematical Sciences, High Point University, 1 University Parkway, High Point, NC, 27268}
\email{agrahams@highpoint.edu}

\author{David McCune}
\address{David McCune, Department of Physics and Mathematics, William Jewell College, 500 College Hill, Liberty, MO, 64068-1896}
\email{mccuned@william.jewell.edu}

\title[Paradoxical Oddities in Two Multiwinner Elections from Scotland]{Paradoxical Oddities in Two Multiwinner Elections from Scotland}

\begin{abstract}
  Ranked-choice voting anomalies such as monotonicity paradoxes have been extensively studied through creating hypothetical examples and generating elections under various models of voter behavior. However, very few real-world examples of such voting paradoxes have been found and analyzed.  We investigate two single-transferable vote elections from Scotland that demonstrate upward monotonicity, downward monotonicity, no-show, and committee size paradoxes. These paradoxes are rarely observed in real-world elections, and this article is the first case study of such paradoxes in multiwinner elections.
\end{abstract}

\maketitle


\section{Introduction}

If you were a candidate in an election, would you prefer more support from voters or less? Put another way: would you prefer that your campaign staff persuade more voters to vote for you, or fewer? These questions seem silly, because of course you would want more support from voters. Surprisingly, when using certain voting methods it is actually possible for more voter support to produce a worse result for a candidate; such an outcome is a type of \emph{monotonicity paradox}. In \cite{GM}, we searched for various types of monotonicity paradoxes in 1,079 single-transferable vote (STV) elections from a database of Scottish local government elections. The purpose of this article is to present, in detail,  two of the most interesting elections revealed by our search. These two elections arguably are the most paradox-riddled real-world political ranked choice elections ever documented, perhaps rivaled only by four single-winner examples from the United States: the 2009 mayoral election in Burlington, Vermont (\cite{FT} and \cite{ON}); a 2021 city council race in Minneapolis, Minnesota \cite{MM};  the 2022 Special Election for US House in Alaska \cite{GM2}; and the 2022 District 4 School Director election in Oakland, CA \cite{Mc}. The first election we present is the 2017 council election in the Buckie Ward of the Moray Council Area, which demonstrated the most extreme instance of a committee size monotonicity paradox ever observed in an actual election. The second election is the 2012 council election in the Ste\`{o}rnabhagh a Tuath Ward of the Na h-Eileanan Siar Council Area, which demonstrated upward and downward monotonicity paradoxes, as well as a no-show paradox. To contextualize these elections, as part of our discussion we indicate how often these kinds of paradoxes occur in Scottish local government elections.



\section{Preliminaries:  Single Transferable Vote and Monotonicity Paradoxes}\label{prelim_section}

We begin with a description of the STV election method that is used in Scottish local government elections, which are almost always multiwinner elections. It is difficult to provide a complete definition of STV in a concise fashion and thus we provide a high level description, accompanied by examples. A complete description of the Scottish STV rules can be found at \url{https://www.opavote.com/methods/scottish-stv-rules}.

Let $n$ denote the number of candidates running in an election and $S$ denote the size of the winner set, which in our context is the number of legislative seats available.  We usually denote the candidates $A$, $B$, $C$, etc., unless we are considering a real-world election, in which case we use actual candidate names. In an STV election, each voter casts a ballot which expresses a preference ranking of the candidates; when describing ballots, we use the symbol $\succ$ to denote that a voter prefers one candidate to another. For example, if $n=4$ then a voter could express that they rank $C$ first, $B$ second, $D$ third, and $A$ last, casting the ballot $C \succ B \succ D \succ A$. In the Scottish elections voters are not required to provide a complete ranking of the $n$ candidates, so a voter could cast the ballot $D \succ B$, for example. Once all ballots are cast they are aggregated into a \emph{preference profile}, which shows how many voters cast each type of ballot. An example of a preference profile with 499 voters and 4 candidates is given in Table \ref{pref_profile}. The number 96 in the second column denotes that 96 voters cast the ballot $A \succ B$; the other numbers communicate similar information about the number of voters who cast the corresponding ballot in that column. We reserve the letter $P$ to denote a preference profile.

\begin{table}[tbh]
\begin{center}
\begin{tabular}{l|cccccccccc}
Number of Voters&96&10&4&128&68&62&4&57&52&18\\
\hline
1st choice &$A$ & $A$ & $B$ & $B$ & $C$ & $C$ & $C$ & $D$ & $D$ & $D$\\
2nd choice& $B$ & $C$& $C$ & $A$ & $A$ & $B$ & $D$ & $A$ & $C$ & $B$\\
3rd choice&        & $D$ & $A$ & $D$ &        & $A$ & $B$ & $C$ & $B$ &\\
4th choice&        & $B$ &        & $C$  &        &        & $A$  &       &         &\\
\end{tabular}

\end{center}
\caption{An example of a preference profile $P$ with 4 candidates and 499 voters.}
\label{pref_profile}
\end{table}

The STV method takes as input a pair $(P,S)$, which we call an \emph{election}, and outputs a set of winners $W(P,S)$ of size $S$ (we note that we avoid the issue of ties, as they do not appear in our empirical work). To calculate the winner set, STV works as follows: the method proceeds in rounds where in a given round either a candidate is eliminated from the election because they received too few first-place votes, or a candidate is elected to fill one of the $S$ seats because they received enough first-place votes. To earn a seat, a candidate's first-place vote total must surpass the election's \emph{quota}, which is defined by \[\text{quota } = \left\lfloor \frac{\text{Number of Voters}}{S+1}\right\rfloor +1.\]

In a given round, if no candidate's first-place vote total achieves the  quota  then the candidate with the fewest first-place votes is eliminated, and this candidate's votes are transferred to the next candidate on their ballots who has not been elected or eliminated. If a candidate's first-place vote total is greater than or equal to the quota, that candidate is elected and the votes they receive above quota (the candidate's \emph{surplus votes}) are transferred proportionally to the next non-eliminated and non-elected candidate which appears on the ballots being transferred.  In effect, the first-place votes that $A$ has earned in order to reach quota are ``locked in'' for $A$, and only $A$'s surplus votes are transferred to other candidates. We fully illustrate this proportional transfer process in the examples below.

The method continues in this fashion until $S$ candidates are elected, or until some number $S'<S$ of candidates have been elected by surpassing quota and there are only $S-S'$ candidates remaining who have not been elected or eliminated.  To illustrate the STV process we calculate $W(P,1)$ and $W(P,2)$ for the preference profile $P$ from Table \ref{pref_profile}.

When $S=1$ the quota is $\lfloor499/2\rfloor +1=250$ and a candidate must earn a majority of first-place votes to win. $A$ receives 106 first-place votes, $B$ receives 132, $C$ receives 134, and $D$ receives 127; no candidate achieves a majority and thus $A$, the candidate with the fewest first-place votes, is eliminated. As a result, 96 votes are transferred to $B$ and 10 votes are transferred to $C$, meaning that now $B$ has earned 228 first-place votes and $C$ has earned 144. Candidate $D$ receives no transfers from the elimination of $A$, and thus $D$ is eliminated because of their 127 first-place votes. As a result, 109 votes are transferred to $C$ (note that the 57 votes from the $D \succ A \succ C$ column are transferred to $C$ because $A$ was previously eliminated) and 18 are transferred to $B$, resulting in a victory for $C$ because they have earned 253 votes. Thus, $W(P,1)=\{C\}$. 

We note that in actual elections it is almost never possible to nicely display the preference profile because there are too many different types of ballots cast. Thus, election offices often display how the election unfolded using what we term a \emph{votes by round table}, where the number of first-place votes for each candidate in each round is displayed. Table \ref{first_votes_by_round} shows this table for $(P,1)$, where the bold font denotes that a candidate is elected. For the real-world elections we discuss in subsequent sections we display only this kind of table, as the preference profiles are prohibitively large. 

\begin{table}[tbh]
\begin{center}

\begin{tabular}{ccc}

\begin{tabular}{c|c|c|c}

\multicolumn{4}{c}{$S=1$, $\text{quota}=250$}\\
\hline
\hline
Candidate& \multicolumn{3}{c}{Votes By Round}\\
\hline
$A$& 106&&\\
$B$& 132&228&246\\
$C$& 134& 144& \textbf{253}\\
 $D$& 127&127&\\
\hline
\end{tabular}
&
&
\begin{tabular}{c|c|c|c|c}

\multicolumn{5}{c}{$S=2$, $\text{quota}=167$}\\
\hline
\hline
Candidate& \multicolumn{4}{c}{Votes By Round}\\
\hline
$A$& 106&&&\\
$B$& 132&\textbf{228}&&\\
$C$& 134& 144& 145.07&\\
 $D$& 127&127&161.25&\textbf{175.25}\\
\hline
\end{tabular}

\end{tabular}
\end{center}
\caption{The votes by round tables for the elections $(P,1)$ (left) and $(P,2)$ (right), where $P$ is the preference profile from Table \ref{pref_profile}.}
\label{first_votes_by_round}
\end{table}

When $S=2$ the quota is $\lfloor499/3\rfloor +1=167$ and a candidate must receive slightly more than a third of the first-place votes to earn one of the two available seats (or, if not enough candidates achieve quota, a candidate could be elected by surviving to the end of the elimination-and-transfer process). No candidate achieves quota in the first round, and thus $A$ is eliminated and their votes transferred, as with the $S=1$ election. In this case, however, after votes are transferred $B$ has earned enough first-place votes to surpass quota, receiving $228-167=61$ surplus votes. These votes must now be transferred proportionally to the next (non-eliminated and non-elected) candidates on the ballots with $B$ at the top.  Candidate $D$ receives $128/228 = 56.14\%$ of $B$'s surplus vote,  $C$ receives only $4/228=1.75\%$, and the remaining 42.11\% of $B$'s surplus votes, corresponding to voters who originally cast the ballot $A\succ B$, are transferred to no one. As a result  $0.5614 \times 61 =34.25$ votes transfer to $D$, $0.0175 \times 61 =1.07$ votes transfer to $C$, and $0.4211 \times 61=25.69$ votes are dropped from the election. See the right votes by round table of Table \ref{first_votes_by_round}.

At this stage there are only two candidates remaining, neither of whom have achieved quota. Even though it is clear that $D$ wins the final seat, the formal STV algorithm eliminates $C$ and 14 votes are transferred to $D$, resulting in $D$ achieving quota with 175.25 first-place votes (note that $D$ would be elected at this final stage even if they do not achieve quota). Thus, $W(P,2)=\{B,D\}$.

 The STV method outlined in the previous examples, up to some technical details that we omit, has been used for local Scottish elections since 2007. For the purposes of local government, Scotland is partitioned into 32 council areas, each of which is governed by a council. The councils provide a range of public services that are typically associated with local governments, such as waste management, education, and building and maintaining roads. The council area is divided into wards, each of which elects a set number of councilors to represent the ward on the council. Once every five years each ward holds an election in which all seats available in the ward are filled using the method of STV. For \cite{GM} we obtained preference profiles from almost every ward for the elections years 2012, 2017, and 2022 (vote data from 2007 is difficult to obtain). In addition, the councils sometimes hold by-elections when a councilor dies or resigns; we collected preference profiles for as many of these elections as we could. This data collection allowed us to construct a database of 1,079 STV elections for which we have complete vote data in the form of preference profiles; 1,049 of the elections are multiwinner such that $S\ge 2$, and the remaining 30 are single-winner. The two elections we present in this article are outliers in that database with respect to paradoxical strangeness. We conclude this section by defining the paradoxes with which we are concerned.

To motivate the first kind of paradox, the reader may have noticed from our running example that STV can produce surprising outcomes when changing the value of $S$. As shown above, $C$ wins the election $(P,1)$ but is not a winner of the election $(P,2)$. This seems like a strange outcome: if $C$ is the overall ``best candidate,'' as evidenced by winning the single-winner election, why would $C$ not also be a winner when there are two seats to fill? How can $C$ simultaneously be the ``best'' single candidate, but also not be a member of the ``top half'' of the candidates? Following language from \cite{EFSS}, we call this kind of paradoxical outcome a \emph{committee size paradox}, defined formally below.  We note, as discussed in \cite{EFSS}, there is some debate about whether committee size paradoxes are actually undesirable, but they certainly are strange and it is understandable if the affected candidate(s) feel unfairly treated (this kind of paradox is also interesting because of its conceptual relationship to the famous \emph{Alabama paradox} in apportionment theory; see \cite{BY}).

\begin{mydef} 
An election $(P, S)$ is said to demonstrate (or exhibit) a \emph{committee size paradox} if $W(P,S') \not\subset W(P,S)$ for some $1\le S' < S$.
\end{mydef}

To motivate our next kind of paradox, suppose $S=2$ and candidate $B$ does some last-minute campaigning which causes the 18 voters who voted $D \succ B$ in the original election $(P,2)$ to change their minds and vote $B \succ D$ instead; furthermore, four of the voters who voted $D \succ C \succ B$ decide to cast the ballot $B\succ D \succ C$ instead. Otherwise all other voters cast the same ballot they did in the original election in Table \ref{pref_profile}; call the resulting preference profile $P'$. What effect should these ballot changes have on candidate $B$, who (as we saw previously) is a winner of the election $(P,2)$? A sensible answer is that these changes should have no effect on $B$: since $B$ was a winner in the original election and $B$ receives more voter support in moving from $P$ to $P'$, it should be the case that $B \in W(P', 2)$. However, as illustrated in Table \ref{modified_profile}, gaining this extra support would actually cause $B$ not to receive a seat. The votes by round table in the bottom of Table \ref{modified_profile} illustrates how this occurs: $B$ has larger vote totals initially in $P'$, but this extra support changes the order in which candidates are eliminated and elected, so that in the election $(P', 2)$ candidate $D$ is eliminated first, instead of $A$ (as occurred in the original election). Surprisingly, $B\notin W(P',2)=\{A,C\}$. 

\begin{table}[tbh]
\begin{center}

\begin{tabular}{l|ccccccccccc}
Number of Voters&96&10&4&128&68&62&4&57&48&4&18\\
\hline
1st choice &$A$ & $A$ & $B$ & $B$ & $C$ & $C$ & $C$ & $D$ & $D$ &B& $B$\\
2nd choice& $B$ & $C$& $C$ & $A$ & $A$ & $B$ & $D$ & $A$ & $C$ & D&$D$\\
3rd choice&        & $D$ & $A$ & $D$ &        & $A$ & $B$ & $C$ & $B$ & C&\\
4th choice&        & $B$ &        & $C$  &        &        & $A$  &       &         & &\\
\end{tabular}

\vspace{.1 in}

\begin{tabular}{c|c|c|c}

\multicolumn{4}{c}{$S=2$, $\text{quota}=167$}\\
\hline
\hline
Candidate& \multicolumn{3}{c}{Votes By Round}\\
\hline
$A$& 106&163&\textbf{168.60}\\
$B$& 154&154&163.40\\
$C$& 134& \textbf{182}& \\
 $D$& 105&&\\
\hline
\end{tabular}

\end{center}
\caption{ (Top) The preference profile $P'$, obtained by moving $B$ up to the first ranking on 22 ballots from the preference profile $P$ from Table \ref{pref_profile}. (Bottom) The votes by round table for the election $(P', 2)$.}
\label{modified_profile}
\end{table}

When an election has the property that one of the winners becomes a loser by shifting the winner up on some ballots, the election is said to exhibit an upward monotonicity paradox.

\begin{mydef} 
An election $(P,S)$ is said to demonstrate (or exhibit) an \emph{upward monotonicity paradox} if there exists $X\in W(P,S)$ and a set of ballots $\mathcal{B}$ such that moving $X$ to a higher position on the ballots from $\mathcal{B}$, while leaving the relative positions of the other candidates on the ballots unchanged, creates a preference profile $P'$ such that $X \not\in W(P',S)$. 
\end{mydef}

Note that an upward monotonicity paradox can affect only a winning candidate, where that winner becomes a loser by gaining voter support. There is a mirror image paradox which can affect only a losing candidate, where the loser becomes a winner by being moved down on some ballots.

\begin{mydef} 
An election $(P,S)$ is said to demonstrate (or exhibit) a \emph{downward monotonicity paradox} if there exists $X\not\in W(P,S)$ and a set of ballots $\mathcal{B}$ such that moving $X$ to a lower position on the ballots from $\mathcal{B}$, while leaving the relative positions of the other candidates on the ballots unchanged, creates a preference profile $P'$ such that $X \in W(P',S)$. 
\end{mydef}

We leave as an exercise to the reader that the election $(P,2)$ does not demonstrate a downward monotonicity paradox. That is, for either of the losers $A$ and $C$, there does not exist a set of ballots such that moving the given loser down on those ballots creates an election in which the loser becomes a winner. Thus, it is possible for an election to demonstrate an upward paradox but not a downward paradox (and vice versa). To see an example of a downward paradox, note that the election $(P,1)$ does demonstrate such a paradox: the reader can check that if 18 of the voters who cast the ballot $B \succ A \succ D$ cast the ballot $A \succ D \succ B$ instead, the winner of the $S=1$ case changes from $C$ to $B$ because the order in which candidates are eliminated changes. In other words, candidate $B$ loses the election $(P,1)$ because in some sense they receive too much voter support. 

To explain our final kind of monotonicity paradox, suppose that 18 of the voters in the election $(P,2)$ who cast the ballot $C \succ A$ decide not to cast a ballot, and instead abstain from the election. What effect should this have on the electoral outcome? Since $W(P,2)=\{B,D\}$ and these 18 voters support $A$ and $C$, it seems like the abstention of these voters should have no effect on the winner set. However, the reader can check that if these 18 voters are removed from the election, the winner set becomes $\{A,B\}$. That is, when these 18 voters participate in the election neither of their two favourite candidates win, but when they do not participate one of their two favourite candidates wins. These voters would have been better off not voting. This is an example of a \emph{no-show paradox} (sometimes also referred to as an \emph{abstention paradox}), where there exists a set of voters such that removing that set of voters from the election creates a more preferable electoral outcome for them.

The formal definition of a no-show paradox has been defined in different ways for single-winner elections, and very few attempts have been made to create a formal definition in a multiwinner context where voters cast preference ballots. The definition we provide below is the kind of paradox that we searched for in \cite{GM}, but other definitions (either more or less restrictive) are sensible. 

\begin{mydef}
Let $(P, S)$ be an election, and let $X \not \in W(P, S)$ and $Y \in W(P, S)$. Let $\mathcal{B}$ be a set of ballots on which $X$ is ranked higher than $Y$. The election demonstrates (or exhibits) a \emph{no-show paradox} if removing the ballots in $\mathcal{B}$ from $P$ creates a preference profile $P'$ such that $W(P',S)=(W(P, S) -\{Y\}) \cup \{X\}$.
\end{mydef}

Informally, an election exhibits this paradox if there exists a set of voters all of whom prefer a losing candidate $X$ to a winning candidate $Y$, and removing these voters from the election causes $X$ to replace $Y$ in the winner set (but no other changes occur to the winner set).

The election $(P,2)$, where $P$ is the preference profile from Table \ref{pref_profile}, demonstrates three of the four paradoxes we have defined, and thus is quite a paradox-riddled election. Of course, we built this example to demonstrate such paradoxical behavior. When we first started searching the 1,079 Scottish elections for monotonicity paradoxes we wondered if any of the real-world elections demonstrated any paradoxes at all, and if any elections exhibited multiple paradoxes. We now present the two most interesting elections that our search uncovered, the preference profiles for which can be found at \cite{T}.

\section{The 2017 Council Election in the Buckie Ward of the Moray Council Area}\label{moray_election}

\begin{table}[tbh]
\begin{center}

\begin{tabular}{c|c|c|c|c}

\multicolumn{5}{c}{$S=1$, $\text{quota}=1571$}\\
\hline
\hline
Candidate& \multicolumn{4}{c}{Votes By Round}\\
\hline
Cowie& 673&&&\\
Eagle& 1060&1390&1431&\\
 McDonald& 691& 791& 1462&\textbf{1755}\\
 Warren& 716&780&&\\

\hline

\end{tabular}
\vspace{.2 in}

\begin{tabular}{c|c|c|c|c}

\multicolumn{5}{c}{$S=2$, $\text{quota}=1047$}\\
\hline
\hline
Candidate& \multicolumn{4}{c}{Votes By Round}\\
\hline
Cowie& 673&680.27&&\\
Eagle& \textbf{1060}& &&\\
 McDonald& 691& 691.50&849.50&\textbf{1576.22}\\
 Warren& 716&716.58&830.26&\\

\hline

\end{tabular}
\vspace{.2 in}

\begin{tabular}{c|c|c|c|c}

\multicolumn{5}{c}{$S=3$, $\text{quota}=786$}\\
\hline
\hline
Candidate& \multicolumn{4}{c}{Votes By Round}\\
\hline
Cowie& 673&\textbf{826.28}&&\\
Eagle& \textbf{1060}& &&\\
 McDonald& 691& 701.60&710.27& \\
 Warren& 716&728.15&734.36&\textbf{1369.44}\\

\hline

\end{tabular}

\end{center}
\caption{The votes totals by round for the 2017 election in the Buckie Ward of the Moray Council Area for $S \in \{1,2,3\}$. In the actual election, $S=3$.}
\label{Moray_table}
\end{table}

The 2017 election in the Buckie ward of the Moray Council Area was a four-candidate contest between the candidates Tim Eagle, Gordon Cowie, Gordon McDonald, and Sonya Warren. In this election $S=3$, and the bottom of Table \ref{Moray_table} shows how the election unfolded. Eagle achieved quota in the first round, and his election caused enough votes to transfer to Cowie that he achieved quota in the second round, and Warren wins narrowly over McDonald in the final round. The resulting winner set is $\{$Cowie, Eagle, Warren$\}$, leaving McDonald the odd man out. 

The top two tables of Table \ref{Moray_table} show that this election demonstrates a committee size monotonicity paradox: when $S=1$ or $S=2$ McDonald is a winner, but in the actual election he loses. This election is one of only nine in the database of 1,049 Scottish multiwinner elections that exhibit a committee size paradox, and therefore this election represents a very improbable occurrence. What sets this Buckie ward election apart from the other eight elections  which demonstrate this paradox is that it is the only election in the database for which:

\begin{enumerate}
\item the election winner for $S=1$, who is considered the ``strongest'' candidate by STV in the single-winner election case, is not a winner for the actual value of $S$, and
\item the paradox can be observed for an $S$ value smaller than $S-1$. That is, for the eight other elections which exhibited this paradox, if $1 \le S' \le S-2$ then $W(P,S') \subset W(P,S)$.
\end{enumerate}

Gordon McDonald truly is unlucky: he would have won a seat for any $S \in \{1,2,4\}$, but unfortunately for him the actual election contained $S=3$ seats. McDonald is simultaneously the ``best'' single candidate, but also not a member of the ``top three'' (out of four) candidates. That is, according to STV  McDonald is the best candidate when $S=1$, but when $S=3$ he is the worst. Based on the available Scottish STV data, McDonald is the only candidate to suffer such a fate in a multiwinner STV election. As far as we know, this Buckie ward election is the only documented instance of such an extreme example of a committee size paradox in a real-world election. (We note that the definition of a committee size paradox assumes the same underlying vote data for the different numbers of seats. In practice, it is possible that voters would vote differently if the number of seats were different, and thus it is possible that McDonald would not be the $S=1$ winner if there were only one seat. We cannot know how the voters might change their expressed preferences under different numbers of seats, and thus we stick to the original definition of this paradox.) 

This Buckie ward election demonstrates none of the other three monotonicity paradoxes defined in the previous section. In fact, none of the elections from our database which demonstrated a committee size paradox also demonstrated any other kind of monotonicity paradox. There has never been a documented instance of a real-world election demonstrating a committee size paradox which also demonstrates another kind of monotonicity paradox. In that sense, the election $(P,2)$ from Section \ref{prelim_section} represents a purely theoretical possibility.

Whenever a paradox occurs, it is natural to ask: who is ``hurt'' or ``treated unfairly'' by this occurrence? In the case of this election, Gordon McDonald seems to have been ``hurt;'' we cannot blame him if he feels treated unfairly by the workings of the STV method. Voters for whom McDonald is their favourite candidate or voters for whom their preferred winner set is $\{$Cowie, Eagle, McDonald$\}$ perhaps also have a legitimate complaint about the election results. Scottish elections are partisan (that is, each candidate runs as a member of a party or as an independent) and thus we can also ask: was a political party ``hurt'' in this election? If, for example, Eagle and McDonald belonged to party $X$ and they were the only two candidates from that party, it might be unfair that  party $X$ would lose a seat as a result of an increase from $S=2$ to $S=3$. In this case, no party seems to have been hurt: Cowie was an Independent, Eagle was a Conservative, and McDonald and Warren belonged to the Scottish National Party (SNP). Thus, in the increase from $S=2$ to $S=3$ no parties lost any seats; the SNP merely swapped one candidate for another. By contrast, in the election presented in the next section it seems that a political party was hurt by monotonicity paradoxes, as well as a candidate and some voters.

\section{The 2012 Election in the Ste\`{o}rnabhagh a Tuath Ward of the Na h-Eileanan Siar Council Area}\label{eileansiar_election}

Our second election, the 2012  Ste\`{o}rnabhagh a Tuath Ward election of the Na h-Eileanan Siar Council Area, simultaneously demonstrated upward and downward monotonicity paradoxes, as well as a no-show paradox.  In this contest ten candidates competed for four seats; the candidate names are listed in the left column of Table \ref{eliean_siar_original}. In terms of party affiliation, Ahmed and G. Murray belonged to the SNP, Paterson belonged to Labour, and the other seven candidates were Independents. As Table \ref{eliean_siar_original} shows, the winners of the election were MacAulay, R. MacKay, MacKenzie, and G. Murray, resulting in three Independents and one member of the SNP being elected to the council from this ward.

\begin{table}
\begin{tabular}{c|c|c|c|c|c|c|c|c|c}

\multicolumn{10}{c}{$S=4$, $\text{quota}=290$}\\
\hline
\hline
Candidate& \multicolumn{9}{c}{Votes By Round}\\
\hline
Ahmed& 176 & 176.42 & 179.48 & 182.51 & 182.75 & 194.88 & 207.08 & 218.20 & \\
Campbell& 91 & 91.48 & 92.51 & & & & & &\\
MacAulay& 103 & 103.54 & 109.57 & 117.66 & 118.38 & 145.72 & 177.34 & 221.89 & \textbf{249.02}\\
J. MacKay& 89 & 89.78 & 95.81 & 108.05 & 108.81 & 123.24 & & &\\
R. MacKay & 262 & 264.08 & 272.08 & \textbf{294.14} & & & & & \\
 MacKenzie& \textbf{299} &&&&&&&& \\
Morrison& 92 & 93.08 & 97.14 & 104.14 & 104.78 &&&& \\
G. Murray & 174 & 174.66 & 180.66 & 192.69 & 193.24 & 208.40 & 228.67 & 251.11 & \textbf{358.36}\\
M. Murray & 120&121.81&123.81&128.84&129.39&142.57&160.97&&\\
Paterson & 40&40.24 &&&&&&&\\

\hline

\end{tabular}

\caption{The vote by round table for the  2012 election in the Ste\`{o}rnabhagh a Tuath Ward of the Na h-Eileanan Siar Council Area.}
\label{eliean_siar_original}
\end{table}

We begin our analysis of this election by demonstrating an upward monotonicity paradox. The four ballots below were cast in the original election; if these four voters had swapped J. MacKay and MacAulay on their ballots, moving MacAulay up one ranking and ostensibly giving him more support, MacAulay would not have won a seat and Ahmed would have won a seat instead.

J. MacKay $\succ$ MacAulay $\succ$ Paterson $\succ$ R. MacKay

J. MacKay $\succ$ MacAulay $\succ$ R. MacKay $\succ$ Campbell

J. MacKay $\succ$ MacAulay $\succ$ Morrison  $\succ$ MacKenzie

J. MacKay $\succ$ MacAulay $\succ$ M. Murray $\succ$ R. MacKay
\vspace{.1 in}

Table \ref{eliean_siar_upward} shows how this paradoxical outcome occurs: even though MacAulay initially receives more votes after we change the four ballots, shifting him up on these ballots ensures that J. MacKay is eliminated from the election much earlier, causing a ripple effect of other changes that culminate in the election of Ahmed rather than MacAulay. This upward monotonicity paradox occurs in much the same fashion as the paradox for the running example from Section \ref{prelim_section}: a candidate receives more support up front, but that additional support causes a change in the order of elimination or election of other candidates, causing the additional up-front support to eventually be the candidate's downfall.

This example of an upward paradox is particularly interesting (or egregious, depending on your point of view) because of the views expressed by the four voters whose ballots we changed. These voters agree on the following:

\begin{enumerate}
\item Their top two candidates are J. MacKay and MacAulay, one of whom is elected in the actual election.
\item Ahmed is not listed on their ballots, and is not one of their top four candidates. From this we infer that these voters prefer that Ahmed not receive a seat.
\item They do not support the SNP, as neither SNP candidate is listed on their ballots.
\end{enumerate}
It is unfortunate, then, that these voters needed to be very careful about how they ranked their two favourite candidates. The rankings in the original election with J. MacKay at the top achieves a desirable electoral outcome, with one of the voters' top two choices winning a seat. But the election hangs on a razor's edge: if these voters had listed MacAulay as their favourite then a candidate whom they prefer (MacAulay) would have been replaced in the
winner set by a candidate whom they do not prefer (Ahmed), and neither of these voters’ two
favourite candidates would receive a seat. Furthermore, instead of the SNP winning
only one of the four seats, it would have won two, presumably an undesirable outcome for these voters.

Outcomes like this  reinforce arguments made by other researchers that preference ballots should not be used in multiwinner elections (see \cite{BKS} and \cite{R1}, for example). Since $S=4$ and the above voters rank precisely four candidates on their ballots, it is likely these voters are simply expressing their preferred four-person winner set. If the voters wish to communicate that information, it seems troubling that the way the voters rank their top four candidates can have a significant effect on the electoral outcome, and that effect is potentially quite negative and paradoxical from these voters' point of view.

\begin{table}
\begin{tabular}{c|c|c|c|c|c|c|c|c|c}

\multicolumn{10}{c}{$S=4$, $\text{quota}=290$}\\
\hline
\hline
Candidate& \multicolumn{9}{c}{Votes By Round}\\
\hline
Ahmed& 176 & 176.42 & 179.48 & 189.54 & 190.01 & 194.13 & 208.38 & 219.50 & \textbf{246.95}\\
Campbell& 91 & 91.48& 92.51 &99.57 &100.29 & & & &\\
MacAulay& 107 & 107.54 & 113.57 & 125.72 & 127.45 & 138.84 & 172.54 & 216.31 & \\
J. MacKay & 85 & 85.78 & 91.81 &  &  &  & & &\\
R. MacKay & 262 & 264.08 & 272.08 & \textbf{298.14} & & & & & \\
 MacKenzie& \textbf{299} &&&&&&&& \\
Morrison& 92 & 93.08 & 97.14 & 103.17 & 104.46 &121.52&&& \\
G. Murray & 174 & 174.66 & 180.66 & 191.75 & 192.88 & 209.05 & 226.39 & 248.96 & \textbf{284.75}\\
M. Murray & 120&121.81&123.81&134.08&135.24&147.38&163.64&&\\
Paterson & 40&40.24 &&&&&&&\\

\hline

\end{tabular}

\caption{The votes by round table which demonstrates an upward monotonicity paradox.}
\label{eliean_siar_upward}
\end{table}

To demonstrate a downward monotonicity paradox, consider the six ballots from the election displayed below. If we move Ahmed down one ranking on these ballots so that he is ranked second and Campbell first, then Ahmed replaces MacAulay in the winner set.  If only Ahmed had done a small amount of strategically targeted campaigning for Campbell, he could have been a councilor.

\vspace{.1 in}

Ahmed $\succ$ Campbell 

Ahmed $\succ$ Campbell

Ahmed $\succ$ Campbell  $\succ$ J. MacKay $\succ$ R. MacKay

Ahmed $\succ$ Campbell  $\succ$ MacKenzie

Ahmed $\succ$ Campbell  $\succ$ Morrison

Ahmed $\succ$ Campbell  $\succ$ G. Murray $\succ$ J. MacKay $\succ$ R. MacKay  $\succ$ Morrison
\vspace{.1 in}

Table \ref{eliean_siar_downward} shows how this paradoxical outcome occurs: even though Ahmed initially receives fewer votes after we change the six ballots, shifting him down on these ballots ensures that Campbell is eliminated much later than in the original election, causing a ripple effect of other changes that culminate in the election of Ahmed rather than MacAulay, and the SNP would have won two seats instead of one. Once again we see a monotonicity paradox manifest because of a change in the order of elimination or election of other candidates.

This paradoxical occurence has similar troubling implications for the voters involved as the upward paradox. In this case all six voters agree that Ahmed  and Campbell are their two favourite candidates, yet in hindsight we see that the voters had to be very careful about how they chose to rank these two candidates. These six voters are strongly communicating that they wish for both Ahmed and Campbell to represent their ward, and if that's the main information they want to communicate then it seems unfortunate that how they rank their top two candidates can have such profound electoral consequences. Furthermore, note that none of these voters seem to care about MacAulay, yet because they made the ``wrong'' choice at the top of their ballots they caused MacAulay to take a seat away from their favourite candidate Ahmed.

\begin{table}
\begin{tabular}{c|c|c|c|c|c|c|c|c|c}

\multicolumn{10}{c}{$S=4$, $\text{quota}=290$}\\
\hline
\hline
Candidate& \multicolumn{9}{c}{Votes By Round}\\
\hline
Ahmed& 170 & 170.42 & 173.48 & 183.54 & 184.01 & 196.27 & 208.38 & 219.50 & \textbf{246.95}\\
Campbell& 97 & 97.48& 98.51 &105.57 &106.29 & 116.48& & &\\
MacAulay& 103 & 103.54 & 109.57 & 125.72 & 127.45 & 158.14 & 172.54 & 216.31 & \\
J. MacKay & 89 & 89.78 & 95.81 &  &  &  & & &\\
R. MacKay & 262 & 264.08 & 272.08 & \textbf{298.14} & & & & & \\
 MacKenzie& \textbf{299} &&&&&&&& \\
Morrison& 92 & 93.08 & 97.14 & 103.17 & 104.46 &&&& \\
G. Murray & 174 & 174.66 & 180.66 & 191.75 & 192.88 & 207.13 & 226.39 & 248.96 & \textbf{284.75}\\
M. Murray & 120&121.81&123.81&134.08&135.24&149.44&163.64&&\\
Paterson & 40&40.24 &&&&&&&\\

\hline

\end{tabular}

\caption{The votes by round table which demonstrates a downward monotonicity paradox.}
\label{eliean_siar_downward}
\end{table}

Finally, this Ste\`{o}rnabhagh a Tuath Ward election demonstrates a no-show paradox. Consider the four ballots below from the original election.

\vspace{.1 in}

J. MacKay $\succ$ R. MacKay $\succ$ G. Murray $\succ$ Ahmed

J. MacKay $\succ$ R. MacKay $\succ$ G. Murray $\succ$ Ahmed

J. MacKay $\succ$ Ahmed $\succ$ G. Murray $\succ$ R. MacKay

J. MacKay $\succ$ G. Murray $\succ$ Ahmed $\succ$ R. MacKay $\succ$ MacAulay $\succ$ Campbell

\vspace{.1 in}

These voters all agree that their preferred winner set is $\{$Ahmed, J. MacKay, R. MacKay, G. Murray$\}$ and they agree that they prefer Ahmed to MacAulay. The first three voters in particular do not seem to care about MacAulay. Furthermore these voters seem to be supporters of the SNP in some sense, since they rank the two SNP candidates in their top four. 

In the original election, two of the four candidates that these voters want on the council receive seats. If we remove these four voters from the election, in effect making them not participate in the election, the winner set is $\{$Ahmed,  R. MacKay, MacKenzie, G. Murray$\}$. If these voters had abstained from voting then they would have been represented by three of their four favourite candidates, and the SNP receives two seats rather than one. That is, these four voters create a better electoral outcome for themselves by staying home than by voting, and this paradox harms a political party as well as these voters. Table \ref{eliean_siar_noshow} shows how this paradoxical outcome occurs: when these four voters are removed  J. MacKay is eliminated in the third round rather than the sixth (as occurred in the original election), creating round-by-round vote totals similar to those in Table \ref{eliean_siar_downward}. 

\begin{table}
\begin{tabular}{c|c|c|c|c|c|c|c|c|c}

\multicolumn{10}{c}{$S=4$, $\text{quota}=289$}\\
\hline
\hline
Candidate& \multicolumn{9}{c}{Votes By Round}\\
\hline
Ahmed& 176 & 176.47 & 179.44 & 188.60 & 189.03 & 193.15 & 207.41 & 218.54 & \textbf{245.95}\\
Campbell& 91 & 91.54& 92.57 &99.64 &100.29 & & & &\\
MacAulay& 103 & 103.60 & 109.64 & 125.80 & 127.39 & 138.76 & 172.47 & 216.25 & \\
J. MacKay & 85 & 85.70 & 91.90 &  &  &  & & &\\
R. MacKay & 262 & 264.31 & 272.31 & \textbf{296.37} & & & & & \\
 MacKenzie& \textbf{299} &&&&&&&& \\
Morrison& 92 & 93.20 & 97.27 & 103.30 & 104.47 &121.54&&& \\
G. Murray & 174 & 174.74 & 180.74 & 190.84 & 191.81 & 207.99 & 225.33 & 247.91 & \textbf{283.68}\\
M. Murray & 120&122.01&124.01&134.31&135.36&147.51&163.77&&\\
Paterson & 40&40.24 &&&&&&&\\

\hline

\end{tabular}

\caption{The votes by round table  which demonstrates a no-show paradox.}
\label{eliean_siar_noshow}
\end{table}

Across the 1,079 elections in our database, we found an upward monotonicity paradox in 21 of them,  a downward paradox in only seven, and we found only four elections which demonstrated both an upward and downward paradox. (We note that in \cite{GM} we found eight additional elections which demonstrate a weaker form of a downward paradox, for a total of 15 elections that violate downward monotonicity in some sense.) As far as we are aware, these four elections are the first multiwinner real-world elections ever documented which exhibit both kinds of paradox. The only other documented elections which demonstrate both paradoxes are a 2021 single-winner municipal election in Minneapolis, MN (see \cite{MM}), and a 2022 single-winner election for School Director in Oakland, CA (see \cite{Mc}).  Thus these four Scottish elections make up two-thirds of the known real-world examples of this double monotonicity issue.

Finally, we found a no-show paradox in only fifteen elections. The 2012 election from the Ste\`{o}rnabhagh a Tuath Ward is the only election in our database, and is the only documented election in any setting of which we are aware, which simultaneously demonstrates an upward monotonicity paradox, a downward monotonicity paradox, and a no-show paradox.

\section{Conclusion}\label{conc}

From the standpoint of social choice theory, the  elections presented in this article are two of the most interesting real-world elections ever documented. They serve to fully demonstrate the paradoxes that can arise from using a rounds-based iterated voting method like STV. We conclude with two broad remarks.

First, given our work in this article and in \cite{GM}, we could ask the very broad question: what kinds of elections tend to demonstrate monotonicity paradoxes? The unhelpful, tautological answer is ``elections like the two in this article.'' A better answer is that paradoxes seem to manifest in elections that are particularly ``close;'' i.e., elections where the electorate is especially ``conflicted'' in some sense.  To test this hypothesis, we use the notion of a \emph{Condorcet committee} (as defined in \cite{G} and \cite{R2}), a multiwinner generalization of the single-winner election concept of a \emph{Condorcet winner}, as a way of testing if an electorate is especially conflicted. Informally, a subset of candidates $\mathcal{C}$ of size $S$ in an election $(P,S)$ is a Condorcet committee if and only if each candidate in $\mathcal{C}$ is preferred head-to-head by more voters than each candidate outside $\mathcal{C}$; that is, given any pair of candidates $(A,B)$ where $A \in \mathcal{C}$ and $B \not\in\mathcal{C}$, more voters prefer $A$ to $B$ than the reverse. If $S=1$ then the Condorcet winner is the single candidate in the Condorcet committee, assuming such a candidate exists. In a given multiwinner election $(P,S)$, if a Condorcet committee of size $S$ exists then in some sense the overall electorate is clear about which set of $S$ candidates is the ``best'' (although there is no guarantee STV chooses this set). If no Condorcet committee exists then the election can be viewed as very ``close,'' much like a single-winner election without a Condorcet winner is very ``close.'' (Of course, there are other ways to measure the closeness of an election.)

Of the 1,079 elections in our database only 18 did not contain a Condorcet committee of size equal to the number of seats available in the election, and the two elections in this article are among them. This suggests that if we are searching for paradoxes in real-world elections, finding elections without a Condorcet committee is a good place to start. This observation is reinforced by the fact that the only documented single-winner real-world elections which demonstrate both an upward and downward monotonicity paradox did not contain a Condorcet winner (\cite{Mc} and \cite{MM}). More broadly, the idea that close elections are more likely to produce monotonicity paradoxes supports the findings of previous literature such as \cite{M}. 

Second, in our view the existence of these two elections is not a knock-down argument against the use of STV. Some voting theorists think that STV should not be used because of its susceptibility to monotonicity paradoxes; while such a position is reasonable, we do not endorse it.  We would sympathize with candidates Ahmed and McDonald if they were to say that STV should not be used, but the paradoxes that negatively affected them occur rarely in practice and all voting methods seemingly have flaws. These two elections are fascinating because of the paradoxical extremes they display, but they are outliers in the landscape of real-world elections whose preference profiles we can access.

\end{document}